\documentclass[12pt,preprint]{aastex}
\usepackage{natbib}
\usepackage{graphicx}
\begin{document}
\title{Active galactic nuclei in four metal-poor dwarf emission-line galaxies
}
\author{Yuri I. Izotov}
\affil{Main Astronomical Observatory, Ukrainian National Academy of Sciences,
27 Zabolotnoho str., Kyiv 03680, Ukraine}
\email{izotov@mao.kiev.ua}
\and
\author{Trinh X. Thuan}
\affil{Astronomy Department, University of Virginia, P.O. Box 400325, 
Charlottesville, VA 22904-4325}
\email{txt@virginia.edu}
%\slugcomment{Version of 9/7/96}

\begin{abstract}
We present 3.5m Apache Point Observatory\footnote{The Apache Point Observatory 
3.5-meter telescope is owned and operated by the Astrophysical Research Consortium.} 
second-epoch spectra of four low-metallicity emission-line dwarf
galaxies discovered serendipitously in the Data Release 5 of the Sloan 
Digital Sky Survey (SDSS) to have extraordinary large broad H$\alpha$ 
luminosities, ranging from 3 $\times$ 10$^{41}$ to 2 $\times$ 
10$^{42}$ erg s$^{-1}$. 
  The oxygen abundance in these galaxies 
is very low, varying in the range 12 + log O/H = 7.36 -- 7.99.
Such extraordinarily high broad H$\alpha$ luminosities cannot be accounted 
for by massive stars at different stages of their evolution. 
By comparing with the first-epoch SDSS spectra, we find that
the broad H$\alpha$ luminosities have remained constant
over a period of 3 -- 7 years, which probably 
excludes type IIn supernovae as a possible
mechanism of broad emission. The emission most likely comes from
accretion disks around intermediate-mass black holes with lower mass
 limits  
in the range $\sim$ 5$\times$10$^5$ $M_\odot$ -- 3$\times$10$^6$
$M_\odot$. If so,
these four objects form a new class of very low-metallicity AGN that 
have been elusive until now. 
The absence of the strong high-ionization lines [Ne {\sc v}]
$\lambda$3426 and He {\sc ii} $\lambda$4686 can be understood if 
the nonthermal radiation contributes less than $\sim$ 10\% of the 
total ionizing radiation.
\end{abstract}

\keywords{galaxies: abundances --- galaxies: irregular --- 
galaxies: active -- galaxies: ISM --- H {\sc ii} regions --- 
ISM: kinematics and dynamics}

\section{INTRODUCTION}

Active galactic
nuclei (AGN) are thought to be powered by massive black holes at the centers of
galaxies, accreting gas from their surroundings. Observations of
AGN show that they generally possess a high metallicity, varying from
solar to supersolar metallicities \citep{S98,H02}.
While the derived metallicities do depend on the
detailed model assumptions, this appears to be a solid conclusion.
Gas metallicity is known to be strongly correlated with the stellar
mass of the host galaxy \citep{T04}. Since AGN are usually
found in massive, bulge-dominated galaxies that have converted most of
their gas into stars by the present epoch, their gas metallicities are 
generally high. A question
then arises: do low-metallicity AGN exist? If so, can we find them in
low-mass galaxies?  To address these questions, \citet{Gr06}
have searched the Sloan Digital Sky Survey (SDSS) Data
Release 4 (DR4) spectroscopic galaxy sample of over 500,000 objects to
select out $\sim$170,000 emission-line galaxies with high S/N
spectra. They then use diagnostic line ratios to select out 23,000
Seyfert 2s galaxies.  Imposing an upper mass limit of 10$^{10}$
$M_\odot$ to restrict themselves to low-mass galaxies, they are left
with a sample of only $\sim$ 40 AGN, which they found appear to have
metallicities around half that of typical AGN, i.e. having solar or
slightly subsolar values. The same high metallicity range is found in the 
sample 
of low-mass AGN of \citet{G07}. 
Assessing their findings, \citet{Gr06} are led to
another question: ``Why are there no AGN with even lower
metallicities?'' In this paper, we suggest that these low-metallicity AGN 
do exist although they are extremely rare.   
%We discuss a sample of four low-metallicity
%emission-line dwarf galaxies that appear to harbor an AGN and that
%have been discovered serendipitously in the SDSS.

\begin{figure*}
\figurenum{1}
%\epsscale{0.9}
\hbox{\includegraphics[angle=0,width=0.255\linewidth]{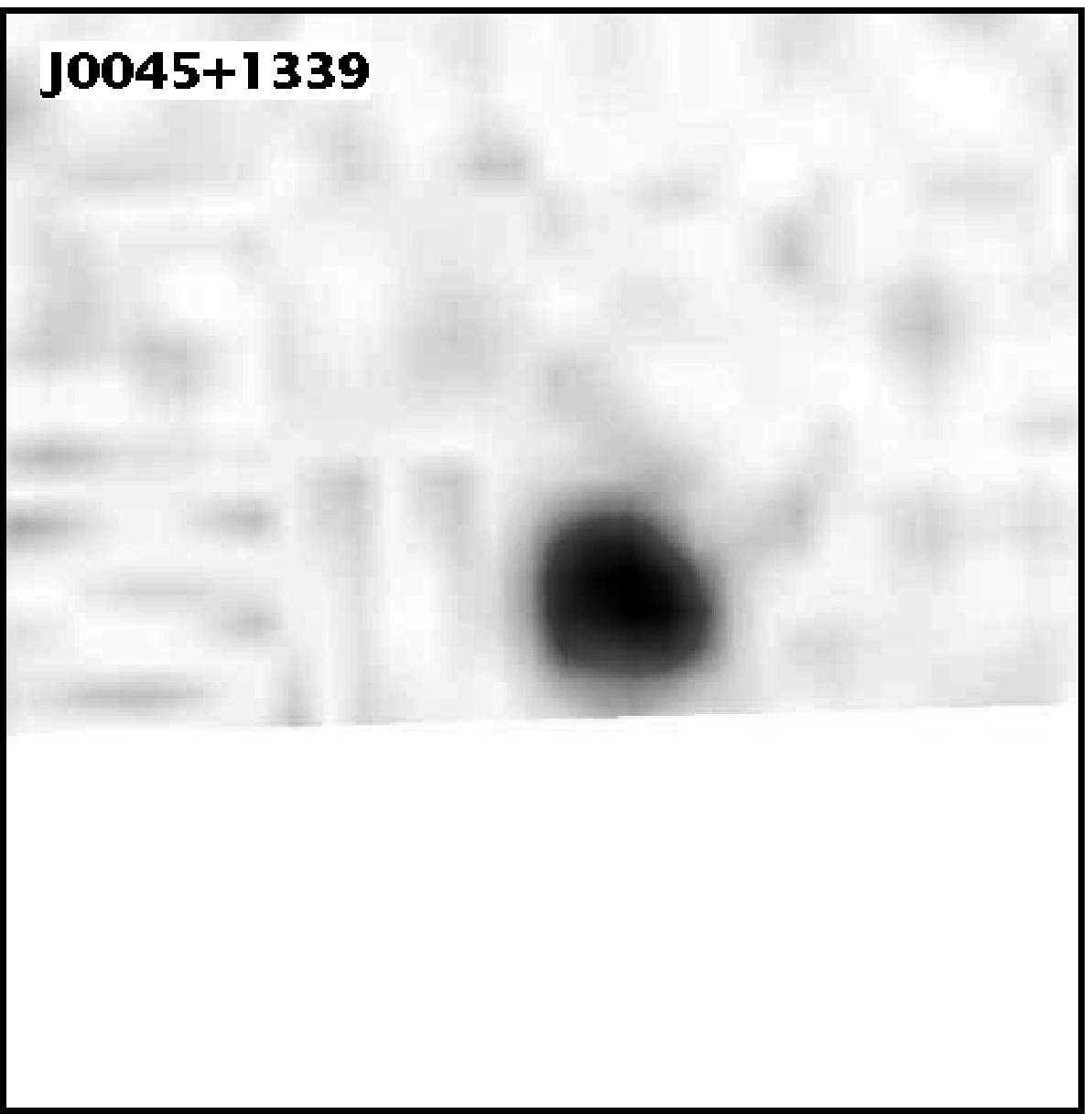} 
\includegraphics[angle=0,width=0.25\linewidth]{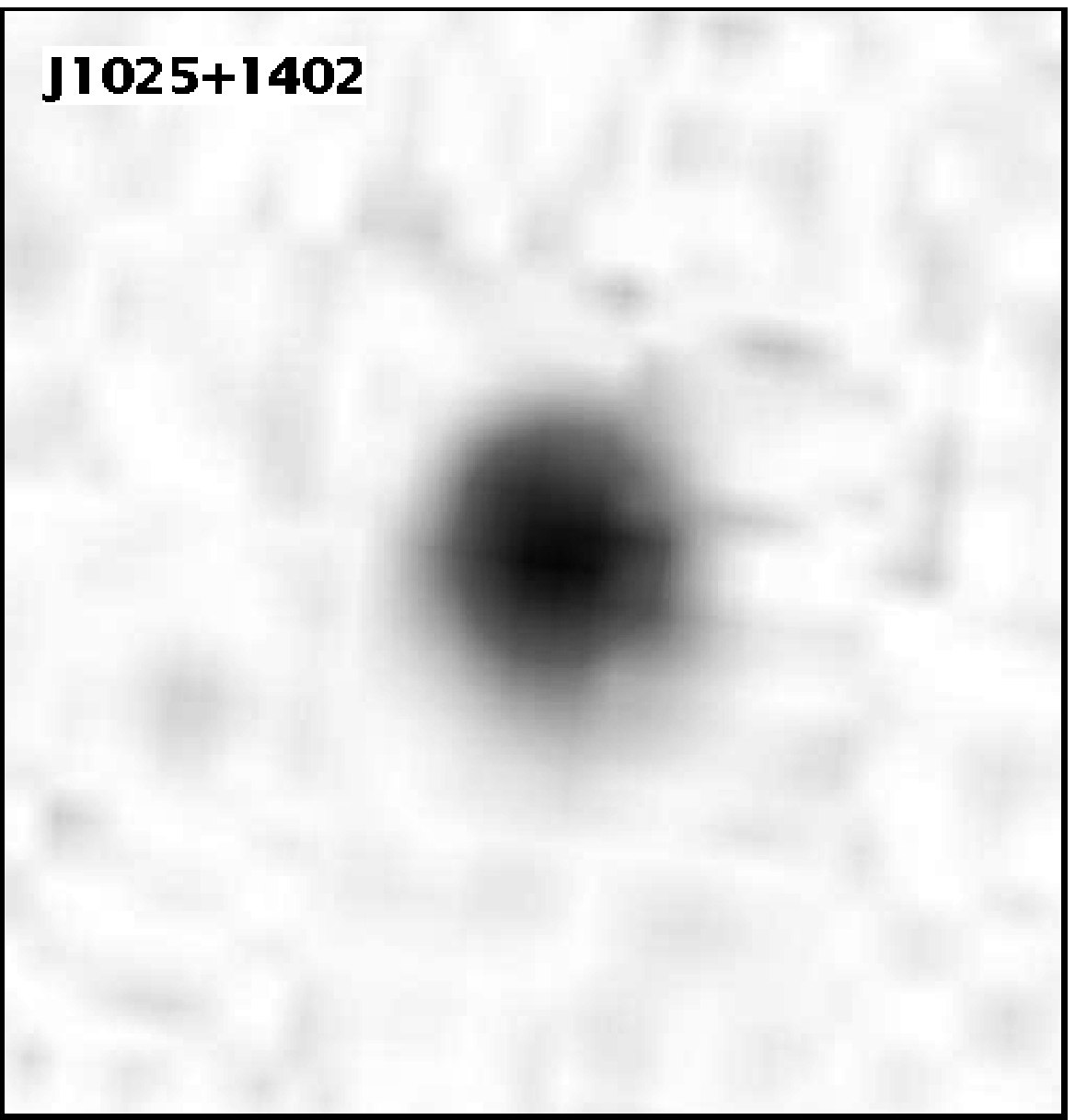} 
\includegraphics[angle=0,width=0.25\linewidth]{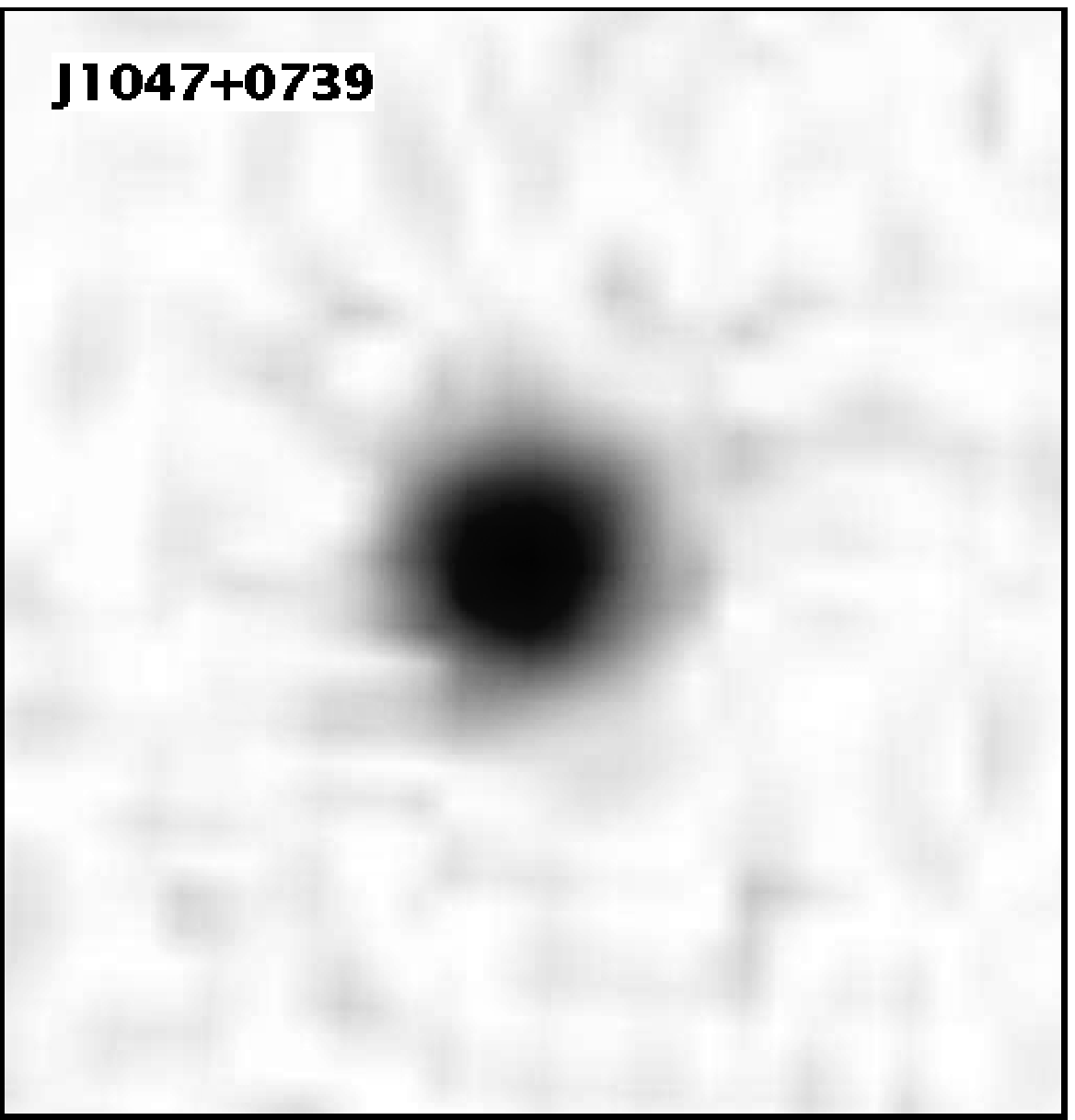} 
\includegraphics[angle=0,width=0.25\linewidth]{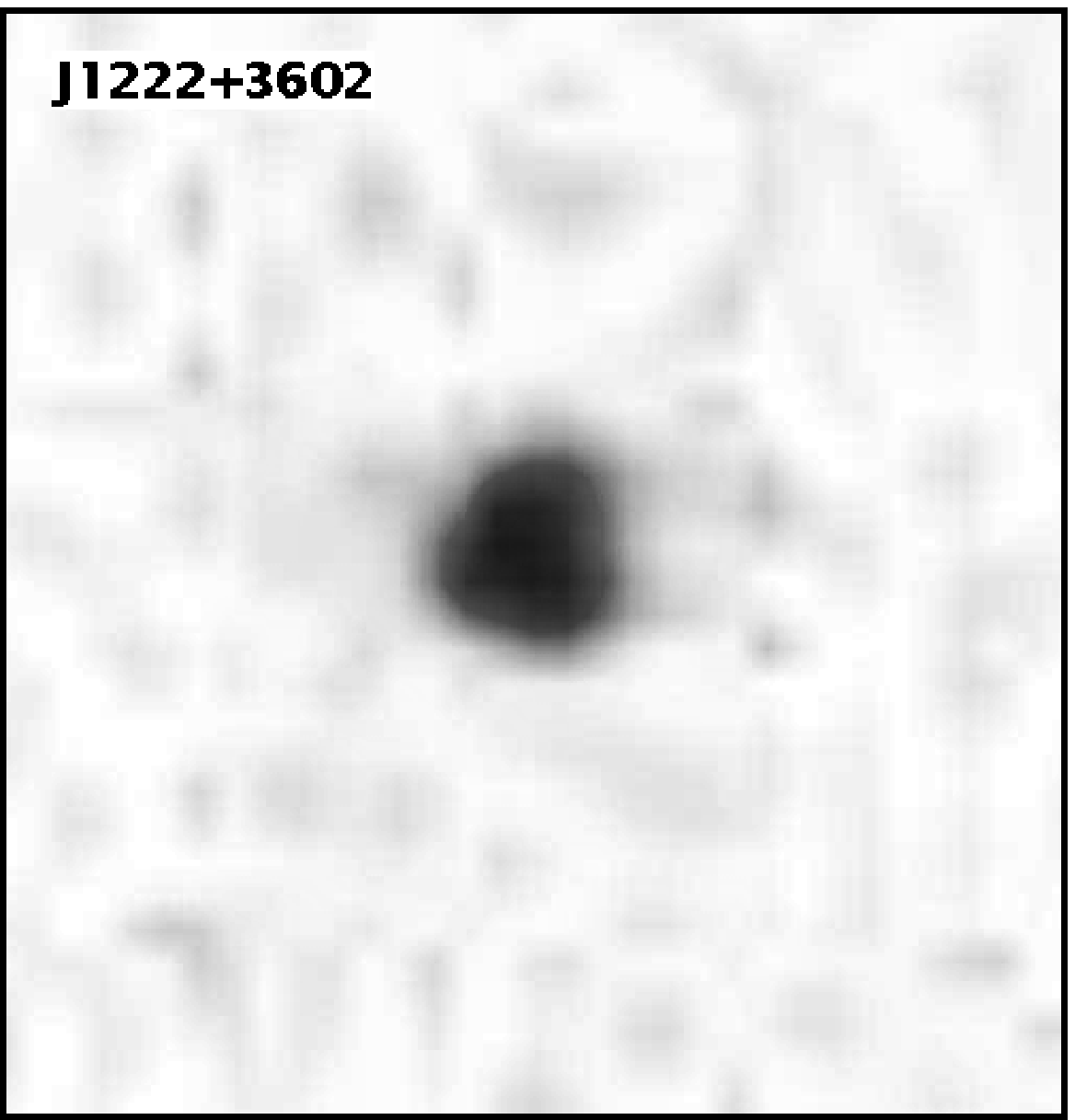} 
}
\figcaption{13\arcsec$\times$13\arcsec\ SDSS images of low-metallicity AGN.
The angular diameters of the objects vary between 1\arcsec\ and 2\arcsec, 
barely larger than the seeing disk.
\label{images}}
\end{figure*}

\begin{figure*}
\figurenum{2}
%\epsscale{0.9}
\hbox{\includegraphics[angle=-90,width=0.8\linewidth]{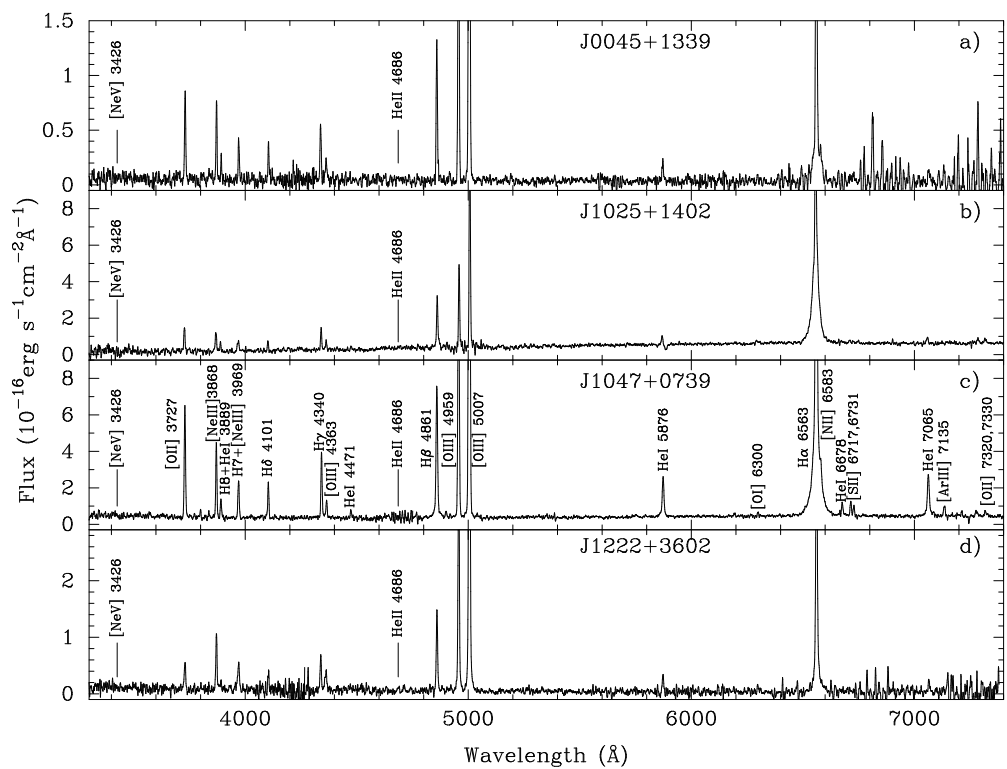} }
%\plotfiddle{spectraAPO.ps}{1pt}{-90.}{300.}{430.}{-10.}{0.}
%\plotone{spectraAPO.ps}
\figcaption{Redshift-corrected 3.5 m Apache Observatory second-epoch 
spectra of four low-metallicy emission-line dwarf galaxies thought to contain
AGN. The locations of the non-detected [Ne {\sc v}] $\lambda$3426 and
He {\sc ii} $\lambda$4686 high-ionization 
emission lines are shown in all panels. Other
emission lines are labeled in panel c). \label{spectra}}
\end{figure*}

In the course of a long-range program to search for
extremely metal-deficient emission-line dwarf galaxies, \citet{I07}
have used the SDSS DR5 database of 675,000
spectra to assemble a large sample of emission-line galaxies. Two
criteria were applied: 1) the [O {\sc iii}] $\lambda$4363 line must be
detected to allow for a direct determination of element abundances;
and 2) obvious high-metallicity AGN spectra are excluded. Thus,
contrary to \citet{Gr06} and \citet{G07}, \citet{I07} 
were not specifically looking for AGN. 
These criteria resulted in a sample of $\sim$10,000 emission-line galaxies
(ELG). While studying that sample to look for ELGs with broad
components in their strong emission lines, \citet{I07} came across four
galaxies with very unusual spectra. The general
characteristics of the four galaxies are given in Table
\ref{tab1}. Their
absolute magnitudes  
are typical of dwarf galaxies. Because of 
their relatively large distance ( $z$ $\sim$ 0.1-0.3) and relatively small 
angular sizes ($\sim$ 1\arcsec-2\arcsec, only slightly larger than 
the seeing disk), their 
SDSS images (Fig. \ref{images}) do not show much details. 
They possess a compact 
structure. Two galaxies, J1025+1402 and J1047+0739, have an approximately 
round shape, while the other two more distant galaxies, 
J0045+1339 and J1222+3602,
have a distorted shape suggestive of mergers. Their colors are not blue 
like the other ELGs, but vary from red to yellow to green.
      Their spectra shown in Fig. \ref{spectra},
resemble those of moderately to very low-metallicity high-excitation H
{\sc ii} regions: their oxygen abundances are in the range 12$+$$\log$
O/H $\sim$7.4--7.9, i.e. their heavy element mass fractions vary from
$Z_\odot$/19 to $Z_\odot$/5 if the solar calibration 12$+$$\log$O/H$=$8.65
of \citet{A05} is adopted. \citet{I07} found that there is however a striking
difference: the strong permitted emission lines, mainly the H$\alpha$
$\lambda$6863 line, show very prominent broad components.  These are
characterized by somewhat unusual properties: 1) their H$\alpha$ full
widths at zero intensity $FWZI$ vary from 102 to 158 \AA,
corresponding to expansion velocities between 2200 and 3500 km
s$^{-1}$; 2) the broad H$\alpha$ luminosities $L_{br}$ are
extraordinarily large, varying from 3$\times$10$^{41}$ to
2$\times$10$^{42}$ erg s$^{-1}$.  This is to be compared with the
range 10$^{37}$--10$^{40}$ erg~s$^{-1}$ found by \citet{I07} for the other
ELGs with broad-line emission.  The ratio of H$\alpha$ flux in the
broad component to that in the narrow component varies from 0.4 to
3.4, as compared to 0.01--0.4 for the other galaxies; 3) the Balmer
lines show a very steep decrement, suggesting collisional excitation
and that the broad emission comes from very dense gas
($N_e$$\gg$10$^{4}$ cm$^{-3}$). Evidently, these galaxies are exceedingly rare, since they constitute only 
4/675,000 or 0.0006\% of our original sample. 

To account for the broad line
emission in these four objects, \citet{I07} have considered various physical
mechanisms: a) Wolf-Rayet (WR) stars; b) stellar winds from Ofp or luminous
blue variable stars; c) single or multiple Supernova (SN) remnants 
propagating in the interstellar medium; 
d) SN bubbles; e) shocks
propagating in the circumstellar envelopes of \hbox{type\,IIn} SNe;
and f) AGN. While mechanisms a-d can account for $L_{br}$ $\sim$ 10$^{36}$ 
to 10$^{40}$ erg s$^{-1}$, they cannot provide for 
luminosities that are 30 to 200 times greater. These very large 
luminosities are more likely associated with SN shocks or AGN. 
\citet{I07} have considered 
\hbox{type\,IIn} SNe because their H$\alpha$ 
luminosities are larger ($\sim$10$^{38}$--10$^{41}$ erg~s$^{-1}$) than those 
of the other SN types, IIp and IIl, and they decrease less rapidly. 

To decide whether type IIn SNe or AGN are responsible for the broad emission 
in these galaxies, monitoring of their spectral features on the
relatively long time scale of several years is necessary. If broad features
are produced by IIn type SNe, then we would expect a decrease in the 
broad line luminosities. No significant temporal 
evolution would be expected in the case of an AGN. Additionally, higher
  signal-to-noise ratio spectra are necessary to put better
  constraints on the presence of the high-ionization [Ne {\sc v}]
  $\lambda$3426 and He {\sc ii} $\lambda$4686 emission lines, 
good indicators of a source of hard non-thermal radiation.
In order to check for temporal evolution, we have obtained
second-epoch spectra of the above 
four galaxies with broad emission, using the 3.5m Apache
Point Observatory (APO) telescope\footnote{In fact, \citet{I07} 
have identified 5 objects as AGN candidates. We have not included here the 
fifth candidate, J2230--0006$\equiv$PHL 293B, because its broad H$\alpha$ 
luminosity is only 8.6$\times$10$^{37}$ erg s$^{-1}$, some 10$^3$ - 10$^4$ 
times lower than those of the other four
AGN candidates. The broad H$\alpha$ luminosity of J2230--0006 has been 
erroneously given 
in Table 8 of \citet{I07} as having 10 times its true value. Given this 
relatively low luminosity and the fact that a new 3.5m APO spectrum of 
J2230--0006
shows that its broad hydrogen lines have a P Cygni 
profile with a blue-shifted absorption, the broad emission 
probably originates from 
a stellar wind rather than from an accretion disk around a AGN. 
Most likely, the broad emission in J2230--0006 is caused by
a strong outburst in a bright luminous blue variable (LBV) star. 
Similar 
broad hydrogen emission has been 
detected recently by \citet{P08} in
the extremely metal-deficient dwarf galaxy DDO 68.}.
We describe the observations in \S2. 
We discuss in \S3 the main properties of the broad emission and show that 
they can be accounted for by low-metallicity intermediate-mass 
AGN. Our conclusions are summarized in \S4.

\section{OBSERVATIONS}

New high signal-to-noise ratio 
optical spectra were obtained for the four galaxies listed in Table \ref{tab1},
using the 3.5 m APO telescope on the nights of 2007 November 15 and
2008 February 6. 
The observations were made with the Dual Imaging Spectrograph (DIS) in the
both the blue and red wavelength ranges. A 1\farcs5$\times$360\arcsec\ slit 
was used. In the blue range, we use the B400 grating 
with a linear dispersion of 
1.83 \AA/pix and a central wavelength of 4400\AA,  while in the red range
we use the R300 grating with a linear dispersion 2.31 \AA/pix and 
a central wavelength of 7500\AA.
The above instrumental set-up gave a spatial scale along the slit of 0\farcs8
pixel$^{-1}$ on the night of 2007 November 15 and of 0\farcs4
pixel$^{-1}$ on the night of 2008 February 6 (the latter scale is 
smaller by a factor of 2 than the previous one because we did not perform
pixel binning for the latter  
spectra),  
a spectral range $\sim$3600 -- 9600\AA\ and a spectral resolution of 7\AA\ 
(FWHM).
The slit was oriented along the parallactic angle and the total exposure 
time was 45 minutes for each galaxy. The observations were broken up into 
3 subexposures to allow for removal of cosmic rays. 
The Kitt Peak IRS spectroscopic standard stars Feige 110 (2007 November 15),
Feige 34 and G191B2B (2008 February 6) were observed for flux
calibration. Spectra of He-Ne-Ar comparison arcs were obtained 
at the beginning of each night for wavelength calibration. 

The data reduction procedures are the same as described in 
\citet{TI05}.
The two-dimensional spectra were bias subtracted and 
flat-field corrected using IRAF.
We then use the IRAF
software routines IDENTIFY, REIDENTIFY, FITCOORD, TRANSFORM to 
perform wavelength
calibration and correct for distortion and tilt for each frame. 
 Night sky subtraction was performed using the routine BACKGROUND. 
The level of
night sky emission was determined from the closest regions to the galaxy 
that are free of galaxian stellar and nebular line emission,
 as well as of emission from foreground and background sources.
A one-dimensional spectrum was then extracted from the two-dimensional 
frame using the APALL routine. An extraction aperture of  
1\farcs5 $\times$ 4\arcsec\ was adopted. Its area
is similar to that of the 3\arcsec\ round aperture used in the
SDSS spectra and allows to gather
$\ga$ 90\% of the light of our compact galaxies. Comparison of the 
line intensities of the SDSS \citep{I07} and 3.5m spectra (Table
\ref{tab2}) shows that they agree
well, validating our choice of the extraction aperture.  
Before extraction, the three 
distinct two-dimensional spectra of each object 
were carefully aligned using the spatial locations of the brightest part in
each spectrum, so that spectra were extracted at the same positions in all
subexposures. We then summed the individual spectra 
from each subexposure after manual removal of the cosmic rays hits. 
%The spectra obtained from each subexposure
%were also checked for cosmic rays hits at the location of strong 
%emission lines, but none were found.

The resulting APO spectra of the four galaxies are shown in Fig. \ref{spectra}. 
A strong broad H$\alpha$ emission line is present in all spectra, very
similar to the one seen in the SDSS 
spectra of the same galaxies obtained 3--7 years
earlier \citep{I07}. The broad components in the H$\beta$ emission line are
considerably weaker, suggesting a steep 
Balmer decrement and hence that the 
broad emission originates in a very dense gas.
In the case of J1047+0739, broad He {\sc i} is also present, as in 
the SDSS spectrum \citep{I07}.

\section{RESULTS AND DISCUSSION}

\subsection{Element abundances}

 We have derived element abundances from the narrow emission line fluxes.
These fluxes have been 
measured using Gaussian fitting with the IRAF SPLOT routine. 
They have been corrected for both extinction, using the reddening curve
of \citep{W58}, and underlying
hydrogen stellar absorption, derived simultaneously by an iterative procedure as
described by \citet{ITL94} and using the observed decrements of the 
narrow hydrogen Balmer H$\delta$ $\lambda$4101, H$\gamma$ $\lambda$4340, 
H$\beta$ $\lambda$4861 and H$\alpha$ $\lambda$6563 lines. It is assumed in
this procedure that hydrogen line emission is produced only by 
spontaneous transitions
in recombination cascades, i.e. we neglect possible collisional excitation.
Such a situation usually holds in low-density H {\sc ii} regions ionized by 
stellar radiation such as those considered here. 
\citet{ITS07} have shown that the use of different reddening curves has
little influence on the extinction-corrected fluxes (relative to the
H$\beta$ flux) in the optical range and modify only slightly the extinction
coefficient. The reason is that the spectra are corrected in such a 
way so that the relative intensities of the extinction-corrected hydrogen
lines correspond to their theoretical values.       
The extinction-corrected fluxes 
100$\times$$I$($\lambda$)/$I$(H$\beta$) of 
the narrow lines for each galaxy, together with the extinction coefficient
$C$(H$\beta$), the equivalent width of the H$\beta$ emission line 
EW(H$\beta$), the H$\beta$ observed flux $F$(H$\beta$) and the 
equivalent widths of the 
underlying hydrogen absorption lines EW(abs) are given in Table \ref{tab2}.
The physical conditions and element abundances of the H {\sc ii} regions
in the four AGN candidates are derived from the narrow line fluxes 
following \citet{I06a}.
The element abundances derived from the 2.5m SDSS spectra \citep{I07}
and from the 3.5 m spectra (this paper) are shown in Table \ref{tab3}.
We find the abundances and the abundance ratios obtained
for the four galaxies from the two sets of observations agree well.
The derived abundances are in the range characteristic of low-metallicity dwarf 
emission-line galaxies \citep{I06a}. 
This implies that the narrow emission lines arise
primarily from  
regions ionized by stellar ionizing radiation, and that any  
contribution of a possible AGN component to the narrow line emission is small.

\subsection{Broad emission and diagnostic diagrams}

Using Gaussian fitting, we have 
measured the fluxes of the broad component of the H$\alpha$ line after 
subtraction of the narrow component. These fluxes are shown in 
Table \ref{tab4}. Comparison of SDSS and APO 
broad H$\alpha$ fluxes shows that they have remained nearly constant  
%within the measurement errors 
(with variations $\leq$ 20\%) over a period of 
$\sim$ 3 -- 7 years. This likely rules out the hypothesis that 
the broad line fluxes are due to type IIn SN because their  
H$\alpha$ fluxes 
should have decreased significantly over this time interval. 
There is one known exception,
SN\,1996cr, where the H$\alpha$ flux has remained constant for $7$~yrs 
\citep{B08}, 
although here the luminosity of H$\alpha$ is only $\sim$ 10$^{38}$ erg~s$^{-1}$. 
Maintaining an H$\alpha$ luminosity of  
10$^{41}$ -- 10$^{42}$ erg~s$^{-1}$ for such durations in our objects
would require nearly the entire energy budget of a SN 
($>$ 10$^{50}$ -- 10$^{51}$ ergs).
%and would be
%interesting in its own right. 
We thus rule out type~IIn SNe.

\begin{figure*}
\figurenum{3}
%\epsscale{0.9}
\hbox{\includegraphics[angle=-90,width=0.5\linewidth]{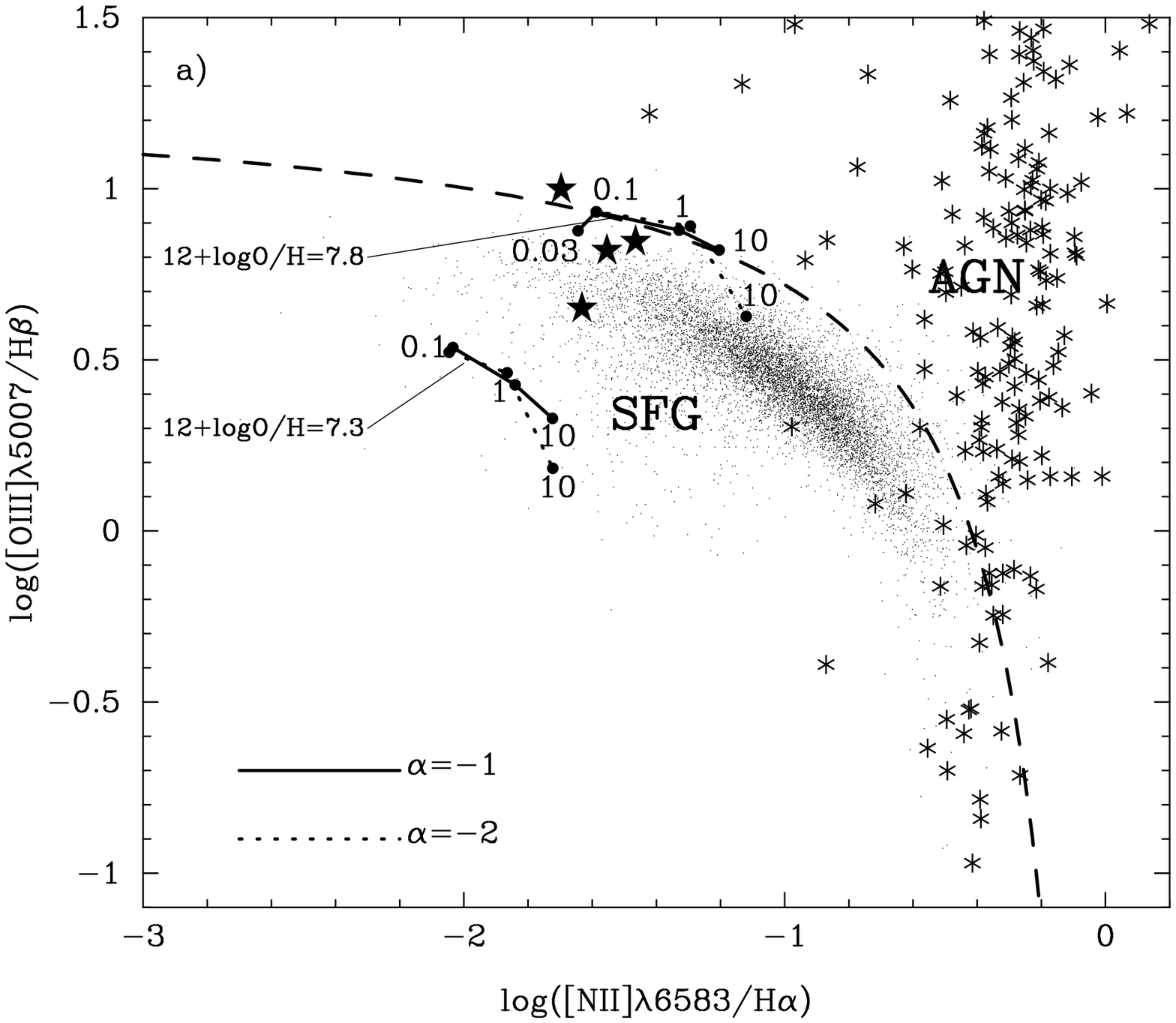} 
\includegraphics[angle=-90,width=0.5\linewidth]{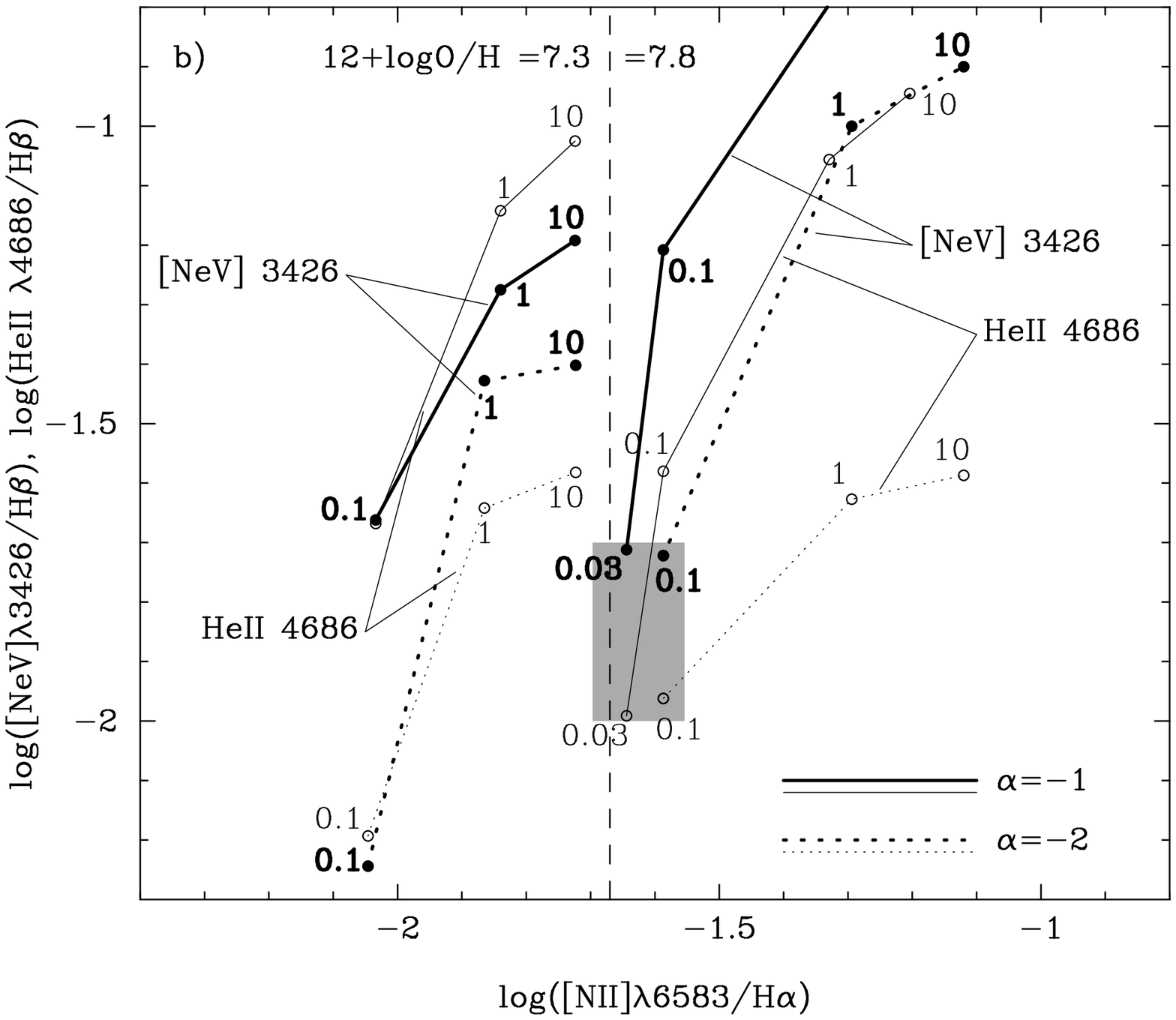} 
}
\figcaption{a) The BPT diagram \citep{B81} for 
low-ionization emission lines. Plotted are the $\sim$10,000 ELGs
  from \citet{I07} (cloud of points), the low-mass black hole sample of 
\citet{G07} (asterisks) and the four low-metallicity AGN in Table \ref{tab1}
  (stars). The dashed line separates star-forming galaxies (SFG) from 
active galactic nuclei (AGN)
  \citep{K03}. The solid and dotted
lines connect CLOUDY photoinization models computed for H {\sc ii}
regions ionized by a composite radiation consisting of different
proportions of stellar and nonthermal radiation.  The two upper curves are
characterized by 12+log O/H = 7.8 and the two lower ones by 12+log O/H =
7.3.  Each model point is labeled by the ratio of
nonthermal-to-thermal ionizing radiation. All curves have been calculated 
adopting a number
of ionizing photons $Q$ = 10$^{53}$ s$^{-1}$ for stellar radiation, 
different slopes of the nonthermal 
spectral energy distributions $f_\nu$ $\propto$ $\nu^\alpha$ 
(solid lines are for $\alpha$ = --1 and dotted lines are for $\alpha$ = --2).
A density $N_e$ = 10$^{4}$ cm$^{-3}$ is adopted. 
Higher densities would shift the curves to the right. 
b) The diagnostic diagram for 
high-ionization emission lines:  
[Ne {\sc v}] $\lambda$3426/H$\beta$ vs. 
[N {\sc ii}] $\lambda$6583/H$\alpha$ (thick lines) and 
He {\sc ii} $\lambda$4686/H$\beta$ vs. 
[N {\sc ii}] $\lambda$6583/H$\alpha$ (thin lines). The same 
CLOUDY models as in a) are shown. The shaded region shows the upper 
intensity limits of
the high-ionization lines [Ne {\sc v}] $\lambda$3426 and 
He {\sc ii} $\lambda$4686 relative to H$\beta$ 
in the four galaxies considered here. The 
dashed vertical line separates models with
12+logO/H = 7.30 (left) from those with 12+logO/H = 7.80 (right).
\label{diagn}}
\end{figure*}

There remains the AGN
scenario.  Can accretion disks around
%intermediate-mass 
black holes 
%(IMBHs, $M$ $\la$ 10$^6$ $M_\odot$) 
in these low-metallicity dwarf galaxies account for their
spectral properties?  The spectra of the four objects do not show
clear evidence for the presence of an intense source of hard
nonthermal radiation: the [Ne {\sc v}] $\lambda$3426, [O {\sc ii}]
$\lambda$3727, He {\sc ii} $\lambda$4686, [O {\sc i}] $\lambda$6300,
[N {\sc ii}] $\lambda$6583, and [S {\sc ii}] $\lambda\lambda$6717,
6731 emission lines, which are usually found in the spectra of AGN,
are weak or not detected. Aside from He {\sc ii} $\lambda$4686, the apparent
weakness of such emission lines, however, may be accounted for by the
low metallicities of our galaxies. None of our objects were detected
in the NVSS or FIRST radio catalogs, demonstrating that they are faint
radio sources.  Another way to check for the presence of an AGN in a
galaxy is to check for its location in the emission-line diagnostic
diagram of \citet{B81} (BPT).  It can be
seen that all four objects lie in the region corresponding to
star-forming galaxies (SFG), to the left of the region occupied by AGN
with low-mass black holes and with metallicities ranging from 2 to 1/4
that of the Sun \citep{G07}. However, their
locations in the SFG region do not necessarily disqualify them as AGN
candidates.  Photoionization models of AGN show that lowering their
metallicity moves them to the left of the BPT diagram, so that they
end up in the SFG region \citep{Gr06,S06}. Thus the BPT
diagram is unable to distinguish between SFGs and low-metallicity AGN.
Admitting that there is an AGN in our dwarf galaxies, can we account
for the weakness of the high-ionization lines?
Photoionization models with only AGN nonthermal ionizing radiation do
predict detectable He {\sc ii} $\lambda$4686 and [Ne {\sc v}]
$\lambda$3426 emission lines.  To make the observed spectra agree with
the models, one solution is to dilute the nonthermal ionizing
radiation from the AGN by thermal radiation from surrounding hot
massive stars.  In Fig. \ref{diagn}a, we show the results of our CLOUDY
calculations \citep{F96,F98} of H {\sc ii} regions ionized by a composite 
radiation consisting of different proportions of stellar and nonthermal
radiation.  Two curves, characterized by different metallicities, are
shown by solid lines: the lower one is for 12$+$$\log$O/H$=$7.3 and the upper 
one is for 12$+$$\log$O/H$=$7.8, typical of the metallicities of our objects.
Each model point is labeled by the ratio $R$ of nonthermal-to-thermal
ionizing radiation. A slope
$\alpha$ = --1 has been adopted for the non-thermal power-law spectrum  
over the whole wavelength range under consideration 
($f_\nu$ $\propto$ $\nu^{\alpha}$). 
The calculations have been done with a number of ionizing photons
$Q_{th}$$=$10$^{53}$~s$^{-1}$ for stellar radiation, $Q_{nonth}$ =
$RQ_{th}$ for nonthermal radiation and
$N_e$$=$10$^{4}$~cm$^{-3}$. Higher densities would move the curves to
the right. For the ionizing stellar radiation, we adopt Costar 
models by \citet{SK97} with  a heavy element mass fraction $Z$ = 0.004 and
an effective temperature of 53,000K corresponding
to a starburst age of $\la$ 3 Myr.
The dotted lines in Fig. \ref{diagn}a 
show the corresponding models with $\alpha$ = --2.
They are very similar to the models with $\alpha$ = --1 when $R$ $\leq$ 1,
but fall below for $R$ $\ga$ 1. 
  It is seen that models with 12$+$$\log$O/H$=$7.8 and in
which the nonthermal ionizing radiation contributes $\la$10\% of the
total ionizing radiation can account well for the location of all four
galaxies in the BPT diagram, independently of the slope of the power-law 
spectrum.

How about the high-ionization lines? Can their absence be due to 
high dust extinction in the central part of the galaxy? 
We consider this possibility unlikely because in this case, 
broad H$\alpha$ emission from the accretion disk surrounding the black hole 
would not be seen also. We turn next to the properties of the ionizing 
spectrum for an explanation.
In Fig. \ref{diagn}b, we show the diagnostic diagram for 
[Ne {\sc v}] $\lambda$3426/H$\beta$ vs. 
[N {\sc ii}] $\lambda$6583/H$\alpha$ (thick lines) and 
He {\sc ii} $\lambda$4686/H$\beta$ vs. 
[N {\sc ii}] $\lambda$6583/H$\alpha$ (thin lines). As in Fig. \ref{diagn}a,
CLOUDY models with $\alpha$ = --1 and --2 are shown by solid and dotted lines.
They are also 
characterized by $Q_{th}$$=$10$^{53}$~s$^{-1}$ for stellar radiation,
 $Q_{nonth}$ = $RQ_{th}$ for nonthermal radiation and
$N_e$$=$10$^{4}$~cm$^{-3}$. Higher densities would shift curves to the right.
The vertical dashed line separates models with 12+logO/H = 7.3 
(Fig. \ref{diagn}b, left) from those with 12+logO/H = 7.8 
(Fig. \ref{diagn}b, right). The shaded rectangle
shows the region of the upper limits of $\sim$ 1\% - 2\% of the H$\beta$ flux, 
set for 
[Ne {\sc v}] $\lambda$3426/H$\beta$ and He {\sc ii} $\lambda$4686/H$\beta$
in our objects, in the observed range of their 
[N {\sc ii}] $\lambda$6583/H$\alpha$ ratio.
There are several points to be made concerning Fig. \ref{diagn}b. 
First, in contrast to 
the low-ionization line fluxes (Fig. \ref{diagn}a),  
the predicted fluxes of the high-ionization
lines are more strongly dependent on the slope of the power-law spectrum of 
the nonthermal radiation.
Second, while there is no significant change of the predicted fluxes
of the He {\sc ii} $\lambda$4686 emission line with metallicity, 
the predicted flux of the [Ne {\sc v}] $\lambda$3426 linearly scales
with 12+logO/H. Third, at a given metallicity, the detectability 
of the high-ionization lines depends on two parameters: one is the 
ratio $R$ of nonthermal-to-thermal ionizing radiation and the other is the 
slope of the power-law ionizing spectrum. If we adopt  
12+logO/H = 7.8 as typical for our galaxies, then Fig. \ref{diagn}b 
shows that models that satisfy the non-detectability limit 
of the high-ionization lines (i.e. that fall within the shaded box) are 
characterized either by a steep slope and a not excessively small 
$R$ ($\alpha$ = --2 and $R$ $\sim$ 0.1) 
or by a shallower slope and a very low $R$ 
($\alpha$ = --1 and $R$ $\sim$ 0.03). 
 We expect intermediate-mass black holes to have accretion disks that 
are hotter than those around supermassive black holes, and hence their 
ionizing spectrum to have a shallower slope, nearer --1 rather than --2.
If that is the case, then the fraction of ionizing nonthermal to thermal 
radiation is very small in our dwarf galaxies, $\sim$ 3\%. 
It is also possible that the absence of strong high-ionization lines is
caused by a 
high covering factor of the accretion disk. In this case the hard 
radiation would be absorbed inside the dense accretion disk 
and no high-ionization forbidden lines would be formed.

\subsection{Black hole virial masses}

We now estimate the masses of the central black holes. 
It has been shown \citep[see e.g.][]{K00} that continuum and 
broad line luminosities in AGN can be used to determine the size and
geometry of the broad emission-line region and the mass of the central
black hole. 
Examining a large sample of broad-line AGN, \citet{G05} 
have found that the H$\alpha$ luminosity 
scales almost linearly with the optical continuum
luminosity and that a strong correlation exists between the 
H$\alpha$ and H$\beta$
line widths. On the basis of these two empirical correlations, those
authors have derived 
the following relations for the central black hole mass:

\begin{equation}
M_{\rm BH}=2.0 \times 10^6 \left(\frac{L_{{\rm H}\alpha}}{10^{42} {\rm ergs\ s^{-1}}}
\right)^{0.55} \left(\frac{{\rm FWHM}_{{\rm H}\alpha}}{10^3{\rm km\ s^{-1}}}
\right)^{2.06} M_\odot \label{eq1},
\end{equation}

\begin{equation}
M_{\rm BH}=4.4 \times 10^6 \left(\frac{L_{5100}}{10^{44} {\rm ergs\ s^{-1}}}
\right)^{0.64} \left(\frac{{\rm FWHM}_{{\rm H}\beta}}{10^3{\rm km\ s^{-1}}}
\right)^2 M_\odot \label{eq2},
\end{equation}
where $L_{{\rm H}\alpha}$ and $L_{5100}$ are respectively 
the broad H$\alpha$ and 
continuum (at $\lambda$ = 5100\AA) luminosities, and FWHM$_{{\rm H}\beta}$ and
FWHM$_{{\rm H}\alpha}$ are respectively the 
full widths at half maximum of the H$\beta$ and
H$\alpha$ emission lines.

In Table \ref{tab5} we list the extinction-corrected broad H$\alpha$ 
luminosities $L$(H$\alpha$)
and continuum luminosities $\lambda$$L_\lambda$(5100) for the four
galaxies, as derived from the SDSS spectra. The extinction
coefficient was set equal to the one derived for the narrow Balmer
hydrogen lines. Since the reddening due to dust extinction in 
dense regions may be larger than that derived from the narrow hydrogen emission
lines, the derived $L$(H$\alpha$) should be considered as lower limits.
The $L$(H$\alpha$) and
$\lambda$$L_\lambda$(5100) of our galaxies follow closely the correlation
between H$\alpha$ and continuum luminosities found by \citet{G05}. This 
implies that our galaxies are very likely the same type of objects as those 
considered by \cite{G05}. Therefore, we can use Eqs. \ref{eq1} -- \ref{eq2}
for the determination of the central black hole masses.
For the flux and luminosity determinations, we first fit the line
profiles by a single Gaussian.
However, we find that single Gaussian fits do not reproduce well 
the broad low-intensity wings
of the H$\alpha$ line. This suggests that that the line broadening is
caused not only by gas motions but also by light scattering. This
hypothesis 
is supported by a steep Balmer decrement (the H$\alpha$-to-H$\beta$ flux
ratio is greater than $\sim$ 7 in our four galaxies),
 implying high gas densities and
possibly high optical depths in the H$\alpha$ line. Therefore, in addition
to a single Gaussian profile fit, we have also fitted the
 H$\alpha$ line by a Voigt
profile which is 
a superposition of both a Gaussian profile and a Lorentzian profile.
In Table \ref{tab5}, we show the FWHMs of the 
Gaussian components in both cases. The FWHMs  
are smaller by a factor of $\la$ 2 in the case of the Voigt profile fitting
as compared to the single Gaussian fitting.   
For the central black hole mass determination, we have adopted the
 Gaussian FWHMs derived from the more accurate Voigt profile fitting.
We have also set FWHM$_{{\rm H}\beta}$ to be equal to FWHM$_{{\rm H}\alpha}$,
following \citet{G05}. The
masses $M_{\rm BH}$(H$\alpha$) and $M_{\rm BH}$(5100) derived from the
broad H$\alpha$ and continuum luminosities are shown in Table \ref{tab5}.
The masses derived from the two methods  are in
good agreement, as expected  
%This is merely a consequence of the correspondence 
from the relation between $L$(H$\alpha$) and $\lambda$$L_\lambda$(5100) for
our objects and the correlation between the H$\alpha$ and continuum 
luminosities found by \citet{G05}. 
The derived masses of the central black holes
in our galaxies are in the range $\sim$ 5$\times$10$^5$ $M_\odot$ -- 
3$\times$10$^6$
$M_\odot$, lower or similar to the mean black hole 
mass of 1.3 $\times$ 10$^6$ $M_\odot$ found by \citet{G07} 
for their sample of low-mass black
holes. Since the luminosities used to derive the masses of the 
central black holes are lower limits, the derived masses 
should also be considered as lower limits.

\section{CONCLUSIONS}

We study here the broad line emission in four 
low-metallicity star-forming dwarf galaxies with 12+logO/H $\sim$ 7.4 -- 8.0,
i.e. with metallicities between 1/19 and 1/5 that of the Sun. 
We have arrived at the following conclusions:

1. The steep Balmer decrements of the broad hydrogen lines and 
the very high luminosities
of the broad H$\alpha$ line in all four galaxies (3$\times$10$^{41}$ to
2$\times$10$^{42}$ erg s$^{-1}$)
suggest that the broad emission arises 
from very dense and high luminosity 
regions such as those associated with 
supernovae (SNe) of type IIn or with accretion 
disks around black holes. However, the relative constancy of the broad 
H$\alpha$ luminosities over a period of 3--7 years likely rules out the SN 
mechanism.
Thus, the emission of broad hydrogen lines is most likely associated  
with accretion disks around black holes.
If so, these four objects would harbor a new class of AGN that are extremely
rare (in 0.0006\% of all galaxies). 
These AGN would be intermediate-mass black holes
residing in low-metallicity dwarf galaxies, with an oxygen abundance that is  
considerably lower than the solar or super-solar metallicity of a typical AGN.

2. There is no obvious spectroscopic evidence for the  
presence of a source of a non-thermal hard ionizing radiation in all   
four galaxies: high-ionization emission lines such as 
He {\sc ii} $\lambda$4686 and [Ne {\sc v}] $\lambda$3426 emission lines
were not detected at the level $\leq$ 1 -- 2 percent
of the H$\beta$ flux. We have calculated a series of CLOUDY models 
with ionizing spectra which include both thermal
stellar and nonthermal power-law ionizing radiation in order to account 
for the absence of the high-ionization lines. 
We find that the predicted fluxes of the high-ionization 
lines are below the detectability level if the spectral energy distribution 
$f_\nu$ $\propto$ $\nu^\alpha$ of the ionizing nonthermal radiation has  
$\alpha$ $\sim$ --1 and the nonthermal ionizing radiation
is significantly diluted by the thermal stellar ionizing radiation
contributing $\la$ 3 percent of the total ionizing radiation, 
or the ionizing spectrum is steeper ($\alpha$ $\sim$ --2), and  
 the nonthermal ionizing radiation
contributes $\la$ 10 percent of the total ionizing radiation.

3. The lower limits of the masses of the central black holes 
$M_{\rm BH}$ of
$\sim$ 5$\times$10$^5$ $M_\odot$ -- 3$\times$10$^6$ $M_\odot$ in our galaxies 
are among the lowest found thus far for AGN.

\acknowledgements
George Privon, George Trammell and David Whelan kindly obtained the 
spectrum of J0045+1339 for us.
We thank Franz Bauer and John Hawley for useful discussions.
Luis Ho and the referee provided very helpful comments on the manuscript.
Y.I.I. is grateful to the staff of the Astronomy Department at the 
University of Virginia for their warm hospitality. 
We thank the financial support of National Science Foundation
grant AST02-05785. The 3.5 APO time was available thanks to a grant from the
Frank Levinson Fund of the Silicon Valley Community Foundation
to the Astronomy Department of the University of Virginia.
    Funding for the Sloan Digital Sky Survey (SDSS) and SDSS-II has been 
provided by the Alfred P. Sloan Foundation, the Participating Institutions, 
the National Science Foundation, the U.S. Department of Energy, the National 
Aeronautics and Space Administration, the Japanese Monbukagakusho, the 
Max Planck Society, and the Higher Education Funding Council for England. 

\clearpage

\clearpage

\begin{deluxetable}{lccccc}
  %\tabletypesize{\tiny}
%  \tabletypesize{\small}
%  \tabletypesize{\footnotesize}
%  \tabletypesize{\scriptsize}
  %\rotate
%  \tablenum{1}
  \tablecolumns{8}
  \tablewidth{0pc}
  \tablecaption{General characteristics of low-metallicity AGN candidates \label{tab1}}
  \tablehead{
\colhead{Object}&\colhead{R.A. (J2000.0)}&\colhead{DEC. (J2000.0)}&\colhead{Redshift}&\colhead{$g$}&\colhead{$M_g$}
}
  \startdata
SDSSJ0045+1339&00 45 29.2&$+$13 39 09&0.29522&21.80&$-$18.56 \\
SDSSJ1025+1402&10 25 30.3&$+$14 02 07&0.10067&20.36&$-$17.66 \\
SDSSJ1047+0739&10 47 55.9&$+$07 39 51&0.16828&19.91&$-$19.23 \\
SDSSJ1222+3602&12 22 45.7&$+$36 02 18&0.30112&21.30&$-$19.10 
  \enddata
  \end{deluxetable}

\clearpage

  \begin{deluxetable}{lrrrrr}
  %\tabletypesize{\tiny}
%  \tabletypesize{\small}
 % \tabletypesize{\scriptsize}
  %\rotate
%  \tablenum{4}
  \tablecolumns{5}
  \tablewidth{0pc}
  \tablecaption{Intensities of narrow emission lines from 3.5m spectra
\label{tab2}}
  \tablehead{
%  \colhead{Ion}&\multicolumn{5}{c}{\sc Intensity}\\       \cline{2-6} 
  \colhead{\sc Ion}
  &\colhead{100$\times$$I$($\lambda$)/$I$(H$\beta$)}
  &\colhead{100$\times$$I$($\lambda$)/$I$(H$\beta$)}
  &\colhead{100$\times$$I$($\lambda$)/$I$(H$\beta$)}
  &\colhead{100$\times$$I$($\lambda$)/$I$(H$\beta$)}}
  \startdata
  & \multicolumn{4}{c}{Galaxy} \\ \cline{2-5}
  &
 \multicolumn{1}{c}{J0045$+$1339 }&
 \multicolumn{1}{c}{J1025$+$1402 }&
 \multicolumn{1}{c}{J1047$+$0739 }&
 \multicolumn{1}{c}{J1222$+$3602 } \\ \tableline
 %  \cline{2-6} 
3727 [O {\sc ii}]                 &  65.45 $\pm$   2.47 &  70.06 $\pm$   6.30 &  94.18 $\pm$   2.20 &  41.36 $\pm$   2.27 \\
3868 [Ne {\sc iii}]               &  50.42 $\pm$   2.05 &  56.01 $\pm$   5.29 &  52.64 $\pm$   1.35 &  64.17 $\pm$   2.68 \\
3889 He {\sc i} + H8              &   \nodata~~~~       &  20.40 $\pm$   4.21 &  18.17 $\pm$   1.02 &  12.74 $\pm$   3.01 \\
3968 [Ne {\sc iii}] + H7          &  29.75 $\pm$   2.51 &  40.16 $\pm$   4.88 &  33.31 $\pm$   1.18 &  47.17 $\pm$   3.50 \\
4026 He {\sc i}                   &   1.18 $\pm$   0.65 &   \nodata~~~~       &   \nodata~~~~       &   \nodata~~~~       \\
4101 H$\delta$                    &  25.04 $\pm$   2.84 &  24.47 $\pm$   4.42 &  26.78 $\pm$   1.10 &  26.95 $\pm$   2.80 \\
4340 H$\gamma$                    &  43.49 $\pm$   2.49 &  50.40 $\pm$   5.80 &  45.90 $\pm$   1.30 &  44.38 $\pm$   4.01 \\
4363 [O {\sc iii}]                &  14.78 $\pm$   1.59 &  24.44 $\pm$   2.98 &  10.68 $\pm$   0.50 &  20.77 $\pm$   1.76 \\
4471 He {\sc i}                   &   \nodata~~~~       &   \nodata~~~~       &   5.03 $\pm$   0.33 &   \nodata~~~~       \\
4861 H$\beta$                     & 100.00 $\pm$   3.12 & 100.00 $\pm$  13.09 & 100.00 $\pm$   2.24 & 100.00 $\pm$   6.33 \\
4959 [O {\sc iii}]                & 244.61 $\pm$   5.94 & 200.48 $\pm$  14.85 & 183.36 $\pm$   3.48 & 283.29 $\pm$   7.73 \\
5007 [O {\sc iii}]                & 747.92 $\pm$  17.18 & 575.78 $\pm$  34.53 & 555.65 $\pm$  10.10 & 853.85 $\pm$  22.21 \\
5876 He {\sc i}                   &   \nodata~~~~       &   \nodata~~~~       &  20.38 $\pm$   0.62 &   \nodata~~~~       \\
6563 H$\alpha$                    & 279.78 $\pm$  13.73 & 271.10 $\pm$  82.71 & 281.42 $\pm$   9.24 & 277.54 $\pm$  14.82 \\
6583 [N {\sc ii}]                 &   6.54 $\pm$   1.03 &   4.06 $\pm$   1.18 &   4.50 $\pm$   0.09 &   \nodata~~~~       \\
6678 He {\sc i}                   &   \nodata~~~~       &   \nodata~~~~       &   6.15 $\pm$   0.27 &   \nodata~~~~       \\
6717 [S {\sc ii}]                 &   \nodata~~~~       &   \nodata~~~~       &   6.53 $\pm$   0.26 &   \nodata~~~~       \\
6731 [S {\sc ii}]                 &   \nodata~~~~       &   \nodata~~~~       &   4.23 $\pm$   0.24 &   \nodata~~~~       \\
7065 He {\sc i}                   &   \nodata~~~~       &   \nodata~~~~       &  28.89 $\pm$   0.74 &   \nodata~~~~       \\
7136 [Ar {\sc iii}]               &   \nodata~~~~       &   \nodata~~~~       &   5.00 $\pm$   0.26 &   \nodata~~~~       \\
 $C$(H$\beta$)                    & 0.135               &  0.110              &  0.400              & 0.100               \\
 EW(H$\beta$) \AA                 & 394                 &    45               &   116               &  196                \\
 $F$(H$\beta$)\tablenotemark{a}   & 0.12                &  0.16               &  0.70               &  0.14               \\
 EW(abs) \AA                      & 5.55                &  0.10               &  5.35               &  5.50               \\
  \enddata
\tablenotetext{a}{in units 10$^{-14}$ erg s$^{-1}$ cm$^{-2}$.}
  \end{deluxetable}

\clearpage

\begin{deluxetable}{lccccccc}
  %\tabletypesize{\tiny}
%  \tabletypesize{\small}
  \tabletypesize{\footnotesize}
%  \tabletypesize{\scriptsize}
  %\rotate
%  \tablenum{1}
  \tablecolumns{8}
  \tablewidth{0pc}
  \tablecaption{Element abundances \label{tab3}}
  \tablehead{
\colhead{Object}&\multicolumn{3}{c}{2.5m SDSS}&\colhead{}
&\multicolumn{3}{c}{3.5m APO} \\ \cline{2-4} \cline{6-8}
\colhead{}&\colhead{12+log O/H}&\colhead{log N/O}&\colhead{log Ne/O}&&
\colhead{12+log O/H}&\colhead{log N/O}&\colhead{log Ne/O}
}
  \startdata
J0045+1339& 7.94 $\pm$ 0.08 & --1.23 $\pm$ 0.17 & --0.77 $\pm$ 0.15 && 7.86 $\pm$ 0.05 & --1.14 $\pm$ 0.08 & --0.83 $\pm$ 0.07 \\
J1025+1402& 7.36 $\pm$ 0.08 & --1.16 $\pm$ 0.30 & --0.86 $\pm$ 0.14 && 7.48 $\pm$ 0.08 & --1.30 $\pm$ 0.14 & --0.74 $\pm$ 0.12 \\
J1047+0739& 7.99 $\pm$ 0.05 & --1.21 $\pm$ 0.09 & --0.73 $\pm$ 0.09 && 7.85 $\pm$ 0.02 & --1.46 $\pm$ 0.03 & --0.68 $\pm$ 0.04 \\
J1222+3602& 7.94 $\pm$ 0.09 & --1.02 $\pm$ 0.24 & --0.84 $\pm$ 0.14 && 7.88 $\pm$ 0.05 & \nodata           & --0.70 $\pm$ 0.07 \\
  \enddata
  \end{deluxetable}

\clearpage

\begin{deluxetable}{lrrrrrr}
  %\tabletypesize{\tiny}
  %\tabletypesize{\small}
%  \tabletypesize{\footnotesize}
%  \tabletypesize{\scriptsize}
  %\rotate
%  \tablenum{1}
  \tablecolumns{7}
  \tablewidth{0pc}
  \tablecaption{Fluxes of the broad H$\alpha$ emission line at 
different epochs \label{tab4}}
  \tablehead{
\colhead{}
&\multicolumn{3}{c}{2.5m SDSS}&\colhead{}
&\multicolumn{2}{c}{3.5m APO} \\ \cline{2-4} \cline{6-7}
  \colhead{Object}
&\colhead{$I_{br}$\tablenotemark{a}}
&\colhead{$I_{br}$/$I_{nar}$\tablenotemark{b}}
&\colhead{Date of Obs.}&&\colhead{$I_{br}$\tablenotemark{b}}
&\colhead{Date of Obs.}
}
  \startdata
J0045+1339& 16.4$\pm$1.7& 41.65&2000 Jan 12&& 18.0$\pm$0.7&2007 Nov 15 \\
J1025+1402&165.0$\pm$5.7&337.49&2004 Mar 11&&192.5$\pm$0.8&2008 Feb 06 \\
J1047+0739&289.2$\pm$9.1&122.34&2003 Jan 31&&224.2$\pm$2.1&2008 Feb 06 \\
J1222+3602& 16.1$\pm$1.8& 71.69&2005 Mar 13&& 22.4$\pm$1.3&2008 Feb 06 \\
  \enddata
\tablenotetext{a}{In units 10$^{-16}$ erg s$^{-1}$ cm$^{-2}$. Flux errors 
are derived taking into account photon statistics in non-flux calibrated 
spectra.}
\tablenotetext{b}{In percent.}
  \end{deluxetable}

\clearpage

\begin{deluxetable}{lrccrrr}
  %\tabletypesize{\tiny}
  \tabletypesize{\small}
%  \tabletypesize{\footnotesize}
%  \tabletypesize{\scriptsize}
  %\rotate
%  \tablenum{1}
  \tablecolumns{7}
  \tablewidth{0pc}
  \tablecaption{H$\alpha$ and continuum luminosities and masses
of the black holes\tablenotemark{a} \label{tab5}}
  \tablehead{
  \colhead{Object}
&\colhead{$L_{br}$(H$\alpha$)\tablenotemark{b}}
&\multicolumn{2}{c}{FWHM$_{br}$(H$\alpha$)}
&\colhead{${\lambda}L_{\lambda}$(5100)\tablenotemark{b}}
&\colhead{$M_{\rm BH}$(H$\alpha$)\tablenotemark{f}}
&\colhead{$M_{\rm BH}$(5100)\tablenotemark{f}} \\ \cline{3-4}
\colhead{}&\colhead{}&\colhead{Gaussian\tablenotemark{c,}\tablenotemark{d}}
&\colhead{Voigt\tablenotemark{c,}\tablenotemark{e}}&\colhead{}&\colhead{}&\colhead{}
}
  \startdata
J0045+1339&2.74$\times$10$^{41}$&1700&1540&7.59$\times$10$^{42}$&2.43$\times$10$^6$&2.00$\times$10$^6$ \\
J1025+1402&3.21$\times$10$^{41}$&1580& 680&3.62$\times$10$^{42}$&5.07$\times$10$^5$&2.50$\times$10$^5$ \\
J1047+0739&1.57$\times$10$^{42}$&1920&1050&2.32$\times$10$^{43}$&3.05$\times$10$^6$&1.91$\times$10$^6$ \\
J1222+3602&2.80$\times$10$^{41}$&1600& 790&7.17$\times$10$^{42}$&6.34$\times$10$^5$&5.10$\times$10$^5$ \\
  \enddata
\tablenotetext{a}{Parameters are derived from the 2.5 m SDSS spectra 
\citep{I07}.}
\tablenotetext{b}{In units erg s$^{-1}$.}
\tablenotetext{c}{In units km s$^{-1}$.}
\tablenotetext{d}{Derived by fitting a Gaussian profile.}
\tablenotetext{e}{Derived by fitting a Voigt profile, a combination of
 Gaussian and Lorentzian profiles. The FWHM is that of the Gaussian profile.}
\tablenotetext{f}{In solar masses.}
  \end{deluxetable}

\clearpage

\clearpage

\clearpage

\end{document}